\documentclass[conference]{IEEEtran}
\IEEEoverridecommandlockouts
\usepackage{cite}
\usepackage{amsmath,amssymb,amsfonts}
\usepackage{algorithmic}
\usepackage{graphicx}
\usepackage{textcomp}
\usepackage{xcolor}
\usepackage{url}
\usepackage{multirow} 
\usepackage{subcaption}
\def\BibTeX{{\rm B\kern-.05em{\sc i\kern-.025em b}\kern-.08em
    T\kern-.1667em\lower.7ex\hbox{E}\kern-.125emX}}
\begin{document}

\title{To Spike or Not to Spike, that is the Question\\

% \author{\IEEEauthorblockN{1\textsuperscript{st} Sanaz M. Takaghaj}
% \IEEEauthorblockA{\textit{Department of Computer Science and Engineering} \\
% \textit{Penn State University}\\
% University Park, USA \\
% sxm788@psu.edu}
% \and
}
%\author{Anonymous authors}
\author{Sanaz M. Takaghaj, Jack Sampson}
\maketitle

\begin{abstract}
Neuromorphic computing has recently gained momentum with the emergence of various neuromorphic processors. As the field advances, there is an increasing focus on developing training methods that can effectively leverage the unique properties of spiking neural networks (SNNs). SNNs emulate the temporal dynamics of biological neurons, making them particularly well-suited for real-time, event-driven processing. To fully harness the potential of SNNs across different neuromorphic platforms, effective training methodologies are essential. In SNNs, learning rules are based on neurons' spiking behavior, that is, if and when spikes are generated due to a neuron's membrane potential exceeding that neuron's spiking threshold, and this spike timing encodes vital information. However, the threshold is generally treated as a hyperparameter, and incorrect selection can lead to neurons that do not spike for large portions of the training process, hindering the effective rate of learning. 

This work focuses on the significance of learning neuron thresholds alongside weights in SNNs. Our results suggest that promoting threshold from a hyperparameter to a trainable parameter effectively addresses the issue of dead neurons during training. This leads to a more robust training algorithm, resulting in improved convergence, increased test accuracy, and a substantial reduction in the number of training epochs required to achieve viable accuracy on spatiotemporal datasets such as NMNIST, DVS128, and Spiking Heidelberg Digits (SHD), with up to 30\% training speed-up and up to 2\% higher accuracy on these datasets.

\end{abstract}


\begin{figure*}[ht]
\centering
\includegraphics[width=\textwidth]{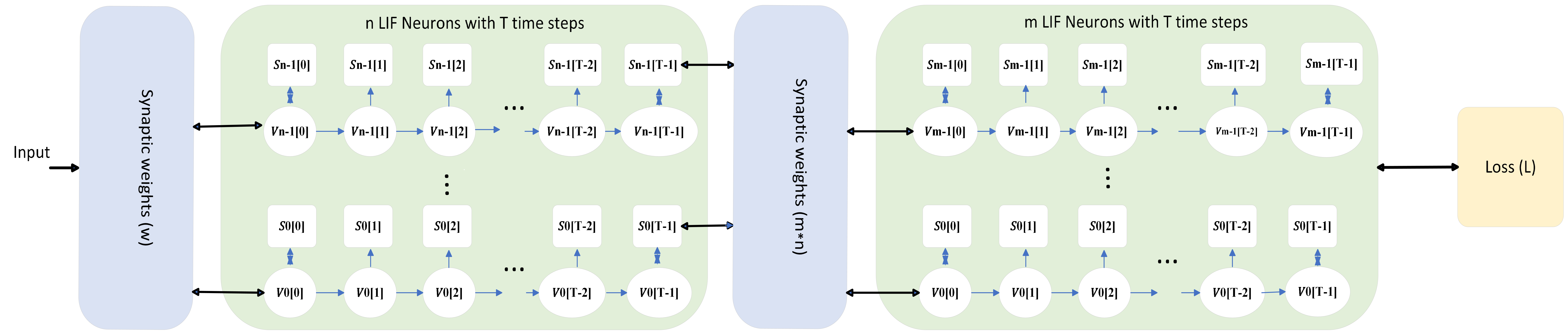}
\caption{Illustration of our (two-layer) SNN architecture with spatiotemporal error backpropagation. Arrows pointing to the right and top indicate the forward path, while arrows pointing to the left and bottom represent the backward path.}\vspace{-8pt}
\label{snn-bptt}
\end{figure*}

\section{Introduction}
Neuromorphic computing is inspired by biological neural networks, which are known for their high energy efficiency in processing stimulus signals and communication. The emergence of DVS cameras, also referred to as ``silicon retina''~\cite{retina} and ``silicon cochlea" devices~\cite{cochlea}, coupled with the use of implementable, dense on-chip artificial synapses~\cite{7047891,sheridan2017sparse, strukov2008missing}, has propelled the field of neuromorphic engineering towards end-to-end event-based models where the data from event-based input devices would be processed via Spiking Neural Networks (SNNs) mapped on neuromorphic hardware fabrics~\cite{TrueNorth-IBM,hoppner2021spinnaker,davies2018loihi,AKIDA,GrAI}. SNNs, in their general form, are composed of Leaky Integrate and Fire (LIF) neurons that have states, and, compared to conventional Artificial Neural Network (ANNs), more closely mimic biological neural networks. LIF neurons are standard neuron models that can be efficiently simulated. SNNs distribute data in the temporal domain rather than the activation amplitude and, when implemented effectively, they will have lower overall memory usage. Beside having state variables (synaptic response current and membrane potential), SNNs also incorporate time as another parameter in their computations. Neuron dynamics and threshold levels specify the timing of spike generation. Since these spikes are sparse, discrete 0/1 events, SNNs are more energy efficient in their computation, and can be implemented in extremely low-power neuromorphic hardware with lower memory access frequency. 
% Many domains, from UAVs to robotics, surveillance and monitoring~\cite{103389, 9138762, 1706816, sandamirskaya2022neuromorphic, tang2019spiking} are poised to benefit from the success of this endeavor.  
  
However, a key roadblock remains for this SNN vision of the future: SNN training is not as mature or reliable as DNN training. Bio-plausible unsupervised training algorithms, such as Hebbian learning and Spike Timing Dependent Plasticity (STDP)~\cite{kempter1999hebbian, song2000competitive}, excel at extracting low-level features but often face challenges with generalization and scalability, limiting their application to shallow SNNs~\cite{diehl2015unsupervised,ferre2018unsupervised}. In reinforcement learning~\cite{seung2003learning, williams1992simple}, synaptic weights are subjected to random perturbations to gauge changes in output error. If the error decreases, the alteration is accepted; otherwise, it is rejected. With large networks, reinforcement learning becomes challenging because the effect of adjusting one weight is overshadowed by noise from others, necessitating numerous trials for meaningful learning.
Due to the remarkable success of error backpropagation in training DNNs, most recent strategies for training SNNs prominently leverage gradient calculations and error backpropagation, achieved through either converting pre-trained ANNs to SNNs~\cite{cao2015spiking, 7280696} or directly training SNNs via error backpropagation~\cite{9556508, perez2021sparse, bohte2000spikeprop, shrestha2018slayer, superspike, lee2016training, kaiser2020synaptic}. 
% The conversion of a pre-trained ANN into an SNN involves a trade-off between accuracy, latency and computational efficiency. And these methods are suited for static datasets where temporal dynamics are absent. 
Directly applying error backpropagation to SNNs is however challenging due to the non-differentiable nature of spiking functions. To address this issue, surrogate gradients~\cite{lee2016training, 8891809, zenke2021remarkable} are used to approximate the spike activation derivative as a continuous function, enabling end-to-end training of SNNs.
% In recent studies, training SNNs has adopted established ANN architectures like VGG, ResNet, and DenseNet~\cite{}. However, there has been relatively less exploration into leveraging the distinctive dynamics inherent to SNNs, which introduce internal memory and a form of nonlinearity less common in ANNs.

% Neurons in SNNs have hyperparameters that control their behavior and dynamics: the firing threshold, membrane time constant ($\tau$), and neuron update time steps.  
% These hyperparameters are bio-inspired, and adjusting them fine-tunes neuron and network behavior. When exposed to input changes (spikes), neurons adapt their membrane potentials in response, while retaining their states in the absence of such stimuli. 
Spiking threshold is a crucial hyperparameter with respect to the activity of a neuron, determining the level of input a neuron must receive before it can generate an output spike. Determining optimal spiking threshold levels that allow neurons to emit enough spikes and, in some cases, precisely timing their emission is a critical task in training SNNs, affecting training convergence rate, inference latency, and accuracy. Increasing the threshold can result in more precise coding of information, with only the most salient features represented by spikes. Decreasing the threshold can lead to more distributed information coding, with more neurons contributing to the representation of each input feature. In backpropagation-based SNN training, achieving the right balance is crucial to avoid the ``dead neuron'' problem. This issue arises when neurons fail to spike during the presentation of an input or input batch, leading to a lack of learning under many learning rules~\cite{bohte2000spikeprop, shrestha2018slayer, lee2016training}. The effectiveness of any SNN training algorithm is therefore sensitive to neuron threshold values, and appropriate threshold values for a given network architecture/dataset are often discovered through offline grid search~\cite{sengupta2019going,7280696}. 

In this study, we investigate how incorporating in-loop threshold learning can reduce the number of dead neurons, leading to faster training and improved inference accuracy. We demonstrate that the adaptive threshold learning can achieve up to a 30\% speed-up in training (measured by the number of epochs before convergence) and up to a 2\% increase in accuracy on neuromorphic datasets NMNIST, DVS128, and Spiking Heidelberg Digits (SHD). The main contribution of this work is an in-depth analysis of the training dynamics and the ``dead neuron'' problem in training SNNs with surrogate gradient-based error backpropagation.

\section{Related work}
\label{sec:motivation}

Several methods have been explored to integrate threshold adjustment into SNNs. The threshold balancing method~\cite{sengupta2019going,7280696} employs a grid search of thresholds to find the best threshold values for ANN to SNN conversion. Another approach involves the use of adaptive spiking neuron models~\cite{yin2020effective, shaban2021adaptive, bellec2018long} and Spike Frequency Adaptation (SFA) mechanisms~\cite{frequency}. These techniques are used to transiently increase the neuron's spiking threshold, thus reducing the neuron's firing rate over time. A comparable strategy is found in the threshold regularization method~\cite{lee2016training}. This approach increases the thresholds of highly active neurons, making them less responsive to input stimuli, while allowing less active neurons to react more easily to subsequent stimuli. The threshold annealing technique~\cite{annealing} increases the threshold level over time, using a small threshold in early epochs and a larger threshold in later epochs. Additionally, some methods like activity regularization~\cite{perez2021sparse, zenke2021remarkable} have proposed lower bound and higher bound values on the neuronal spike counts to control spiking activity levels in SNNs. More recent approaches focus on threshold optimization~\cite{hasssan2024spiking, 9556508, wang2022ltmd, yin2020effective}. These techniques utilizes gradient descent to find the threshold values that minimize a loss function. While improved performance has been reported, there hasn't been a dedicated study on the dynamics of threshold learning and its impact on SNN training through error backpropagation.

\begin{table*}
\caption{Top-1 Test Accuracy Scores}\vspace{-8pt}
\label{benchmark}
\begin{center}\small
\begin{tabular}{|p{1.75cm}|p{3cm}|p{3.5cm}|p{4.5cm}|}
        \hline
        \textbf {Dataset} & \textbf {Training Method} & \textbf {Architecture} & \textbf {Accuracy} \\
        \hline
        \multirow{2}{*}{NMNIST} 
        & Baseline & 34x34x2-500-500-10& 98.89\% \\
        \cline{2-4}
        & Rouser & 34x34x2-500-500-10 & \textbf{99.21\%} \\
        \hline
        \hline
        \multirow{2}{*}{DVS128} & Baseline & 128x128x2-64-11 & 84.72\% \\
        \cline{2-4}
        & Rouser & 128x128x2-64-11 & \textbf{86.46}\% \\
        \hline
        \hline
        \multirow{2}{*}{SHD} & Baseline & 700-200-200-20 & Failed: Training non-convergent\\
        \cline{2-4}
        % & Sparse BP \cite{perez2021sparse} & Regularization& 700-200-200-20 & 77.5\% \\
        % \cline{2-5}
        & Rouser & 700-200-200-20 & \textbf{78.14}\% \\
        \hline
\end{tabular}\vspace{-8pt}
\end{center}
\end{table*}

\begin{table}
\caption{Hyperparameters}\vspace{-8pt}
\label{hyperparameter}
\begin{center}\small
\begin{tabular}{|c|c|c|}
\hline
\textbf{Symbols} & \textbf{Description} & \textbf{Value} \\
\hline
$Th_{init}$ & Initial threshold & 1.25 \\
% $Th_{init}$ & Initial threshold & 1.25, 0.25 \\
% s & Scaling factor & 1.5, 3 \\
s & Scaling factor & 1.5 \\
$\tau$ & Steepness parameter & 3.75 \\
% $\tau$ & Steepness parameter & 3.75, 3 \\
$lr_{W}$ & Synaptic weight's learning rate & 0.001 \\
$lr_{Th}$ & Thrshold's learning rate & 0.001 \\
- & Current decay & 0.75 \\
- & Voltage decay & 0.97 \\
$V_{Rest}$ & Resting voltage & 0 \\
$L$ & Loss function & MSE \\
- & Optimizer & Adam \\
- & Weight initialization & Kaiming\\ 
\hline
\end{tabular}\vspace{-8pt}
\end{center}
\end{table}

\section{SNN Training using spatiotemporal backpropagation and threshold learning}
\label{sec:design}

This section describes the dynamics of Leaky Integrate-and-Fire (LIF) neurons and their training. A LIF neuron $i$ in layer $l$ has two state variables: a synaptic response current $I^{l}_i[t]$ and a membrane potential $V^{l}_i[t]$. The synaptic response current is generated by a leaky integrator which integrates the incoming spikes from a stimuli $S^{l-1}_{j}[t]$. This current then passes through another leaky integrator to produce the membrane potential $V^{l}_i[t]$. When the membrane potential reaches a spiking threshold level $Th^{l}_i$, the neuron sends out a spike $S^{l}_i[t]$ and then resets to $V_{Rest}$. A unit step function models the spiking function:\vspace{-6pt}

\begin{equation}
S^{l}_i[t] =
\left\{
\begin{array}{ll}
    1, & \mbox{if } V^{l}_i[t] \geq Th^{l}_i \\
    0, & \mbox{otherwise}
\end{array}
\right.
\end{equation}\vspace{-8pt}

The spiking function $S^{l}_i[t]$ is non-differentiable and poses a problem for gradient calculations in error backpropagation. To overcome this issue, a surrogate function is used to approximate the derivative of the spiking function:\vspace{-6pt}

\begin{equation}
\label{surrogate}
\frac{\mathrm{d} S^{l}_i[t]}{\mathrm{d}V^{l}_i[t]} = \frac{s}{\tau} e^{-\frac{\left |V^{l}_i[t]-Th^{l}_i\right |}{\tau}}
\end{equation}
\begin{equation}
\label{surrogate_threshold}
\frac{\mathrm{d} S^{l}_i[t]}{\mathrm{d}Th^{l}_i} = -\frac{s}{\tau} e^{-\frac{\left |V^{l}_i[t]-Th^{l}_i\right |}{\tau}}
\end{equation}
\vspace{-6pt}

where $s$ is a scaling factor, and $\tau$ is the steepness parameter ($\tau \rightarrow 0$). Each neuron is composed of synapses with associated synaptic weights $w_{ji}$ and adjustable spiking thresholds $Th^{l}_{i}$. 
% Establishing connections between these LIF neurons constructs a network. Larger networks have an increased number of neurons within each layer that are connected to the neurons in the subsequent layer, forming a fully connected architecture. 
Fig.~\ref{snn-bptt} illustrates our fully connected SNN topology, featuring only two layers for simplicity. In each layer, the membrane potential of each neuron $V_i[t]$ updates every discrete time step $t \in \left \{ 0, 1, \cdots T-1 \right \}$ for a finite number of time steps $T$. 
$S_i[t]$ denotes the spike generated by the neuron $i$ at time $t$. After the final layer, a loss function calculates the difference between the output spike rates and the target spike rate. 
We first calculate the gradient of $L$ with respect to $V^{l}_i$ and $Th^{l}_i$, and then use the chain rule to find the update rules for the synaptic weights $w_{ji}$ and the thresholds $Th^{l}_i$ respectively:\vspace{-6pt}

\begin{equation}
w_{ji} = w_{ji} - lr_{W} * \frac{\partial L}{\partial V^{l}_i[t]}* \frac{\partial V^{l}_i[t]}{\partial w_{ji}}
\end{equation}
\begin{equation}
Th^{l}_i = Th^{l}_i - lr_{Th} * \frac{\partial L}{\partial Th^{l}_i}
\end{equation}\vspace{-8pt}

The learning rate for synaptic weights is denoted as $lr_{W}$, while the learning rate for thresholds is $lr_{Th}$.

\begin{figure}
\begin{center}
\includegraphics[width=0.35\textwidth]{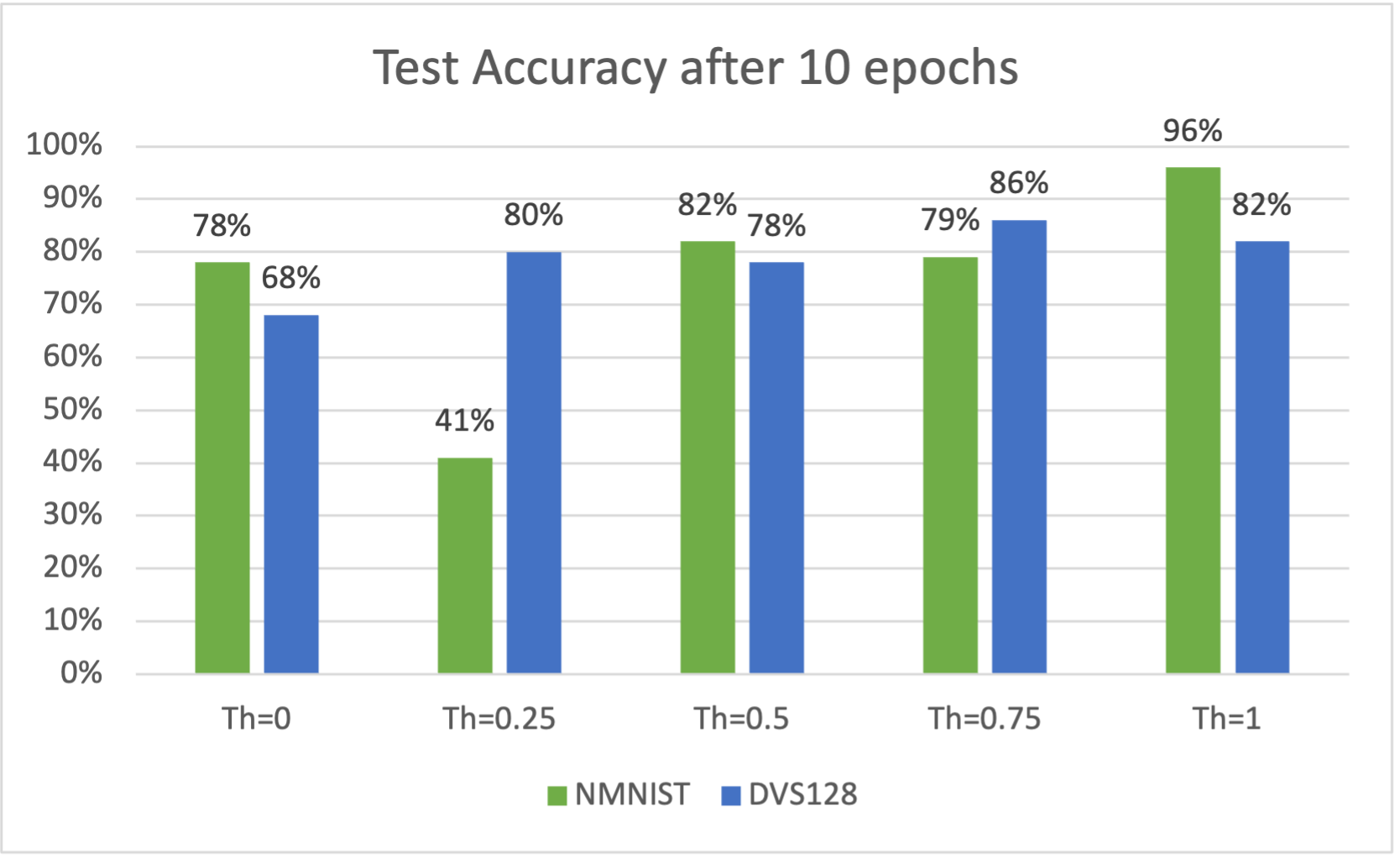}
\label{acc-Th-NMNIST}
\hfill
\end{center}
\caption{Threshold grid search on NMNIST and DVS128. }\vspace{-8pt}
% Among the tested thresholds for NMNIST, a threshold of 1 produces the highest inference accuracy after 10 epochs. For DVS128, this value is 0.75.
\label{acc-10-epochs}
\end{figure}

\section{Experimental Setup}
\label{sec:evaluation}
We used SLAYER from the Lava-DL library~\cite{LAVA} as the baseline for training SNNs with error backpropagation. We then extended SLAYER by incorporating threshold learning through error backpropagation (Rouser). Our evaluation is based on the NMNIST~\cite{orchard}, DVS128~\cite{amir2017low}, and Spiking Heidelberg Digits (SHD)~\cite{cramer2020heidelberg} datasets. 

The NMNIST dataset is a spiking version of the MNIST dataset, where spikes are generated by displaying images on a neuromorphic vision camera equipped with ATIS (Asynchronous Time-based Image Sensor). It contains 60,000 training images, each consisting of 300 time samples, and an additional set of 10,000 test images used to evaluate accuracy. 

The DVS128 dataset is an event-based dataset captured using the Dynamic Vision Sensor (DVS) camera. It comprises recordings of 11 hand gestures and consists of $\approx$20,000-40,000 events per gesture class. The dataset includes 1176 samples with varying temporal dimension allocated for training and 288 samples designated for testing. 

The SHD dataset is an audio-based dataset consisting of spoken digits ranging from zero to nine in both the English and German languages. The audio waveforms have been converted into spike trains using an artificial model of the inner ear. The SHD dataset comprises 8,156 training samples and 2,264 test samples with varying time spans.

\section{Evaluation}

To assess the impact of various threshold levels on the training of SNNs through error backpropagation, we first conducted an experiment using both the NMNIST and DVS128 datasets, systematically varying the threshold values for evaluation while keeping all other hyperparameters and the loss function constant. Fig.~\ref{acc-10-epochs} shows that SNN training convergence is highly sensitive to the threshold level, and the optimal threshold value differs between these two datasets. 

% After 10 epochs of training, we obtained a test accuracy of 96\% on the NMNIST dataset by setting a constant spiking threshold of 1. However, when the threshold was set to 0.25, the performance of the SNN significantly deteriorated, resulting in a subpar test accuracy of 41\%. Similarly, for the DVS128 dataset, we observed an accuracy of 86\% when the spiking threshold was set to 0.75. For a threshold of zero, the accuracy dropped to 68\% (Fig.~\ref{acc-10-epochs}). 

Table~\ref{benchmark} shows the top-1 accuracy of Rouser (with threshold learning) in comparison to the baseline (without threshold learning), along with the network architecture utilized in the study. Table~\ref{hyperparameter} presents the hyperparameters that were utilized in the study. In addition to the improvement in accuracy, we observed greater robustness in Rouser, particularly during training on SHD.

\subsection{Rouser's Learning Dynamic}

\begin{figure}[b]
\centering\vspace{-8pt}
\includegraphics[scale=.5]{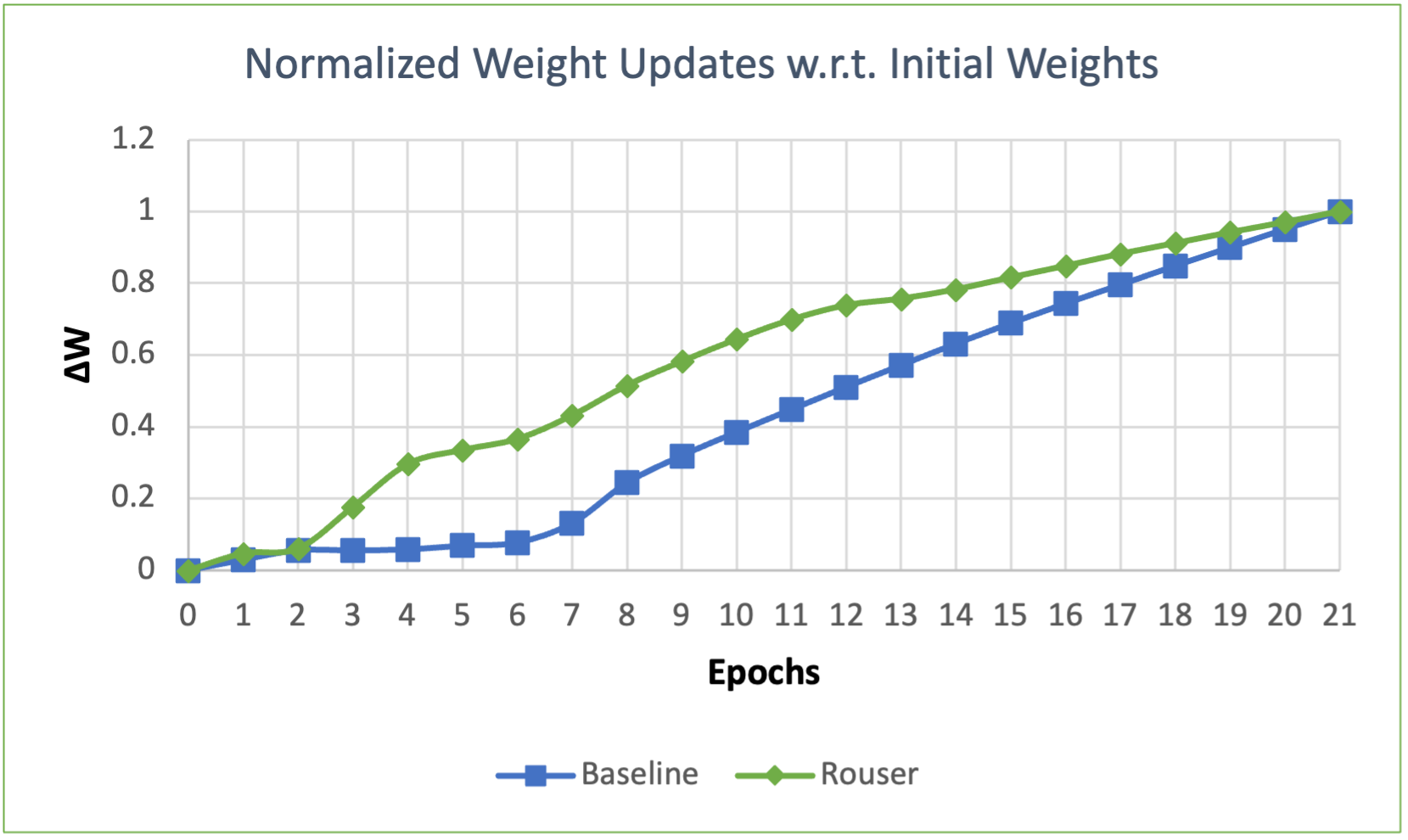}
\caption{Weight updates wrt. initial weights during training.}
\label{WUpdatesInit}
\end{figure}

To better understand the effectiveness of Rouser in SNN training and to gain insights into its convergence latency, we monitored changes in synaptic weights during the training process. Figure~\ref{WUpdatesInit} illustrates these changes over time in comparison to the initial weights. We observed that Rouser exhibited greater weight changes than the baseline, indicating a faster training process.

\begin{figure}
\begin{center}
\begin{subfigure}{0.35\textwidth}
\includegraphics[width=\textwidth]{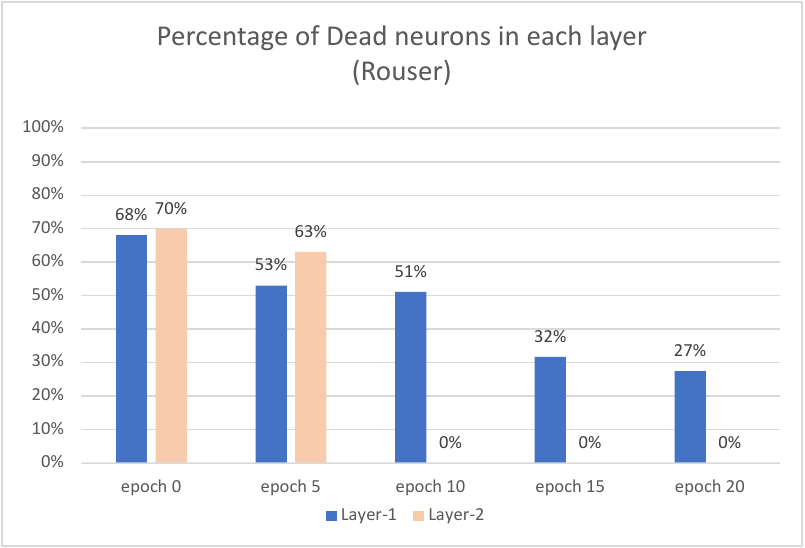}
\caption{}
\label{fig:active-neurons_a}
\end{subfigure}
\hfill
\begin{subfigure}{0.35\textwidth}
\includegraphics[width=\textwidth]{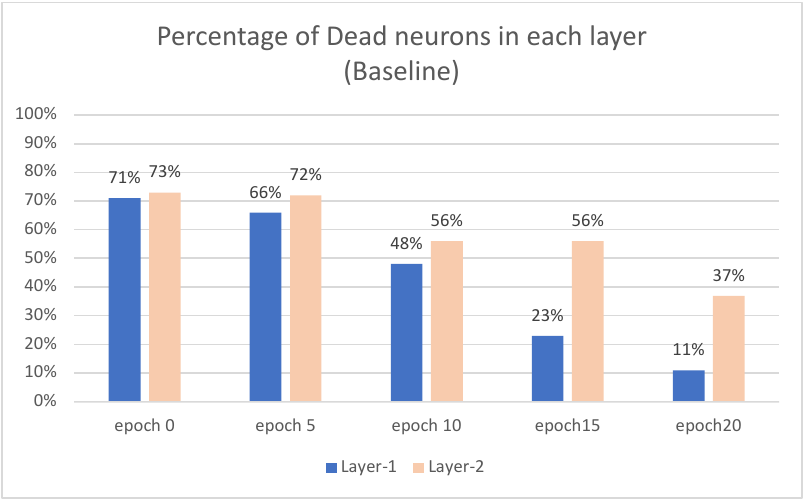}
\caption{}
\label{fig:active-neurons_c}
\end{subfigure}
\end{center}
\vspace{-4pt}\caption{The percentage of ``dead neurons" for each layer during training on NMNIST.}\vspace{-8pt}
\label{dead-neurons}
\end{figure}

\begin{figure}[ht]
\begin{center}
\begin{subfigure}{0.35\textwidth}
    \centering
\includegraphics[width=\textwidth]{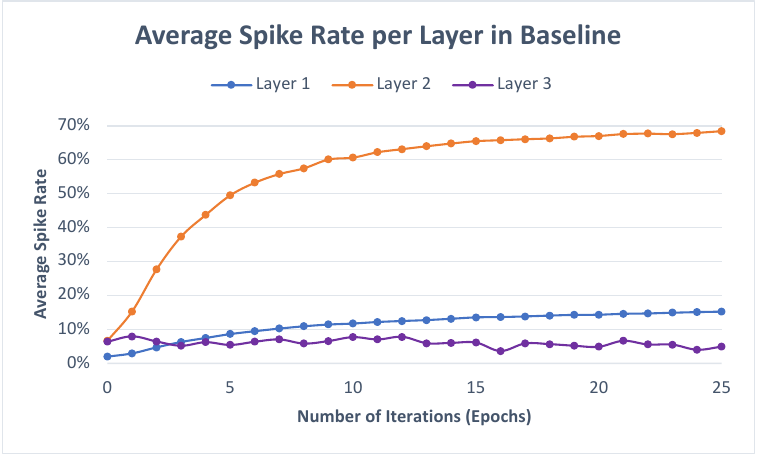}
\caption{}
\label{fig:spikes_a}
\end{subfigure}
\hfill
\begin{subfigure}{0.35\textwidth}
\centering
\includegraphics[width=\textwidth]{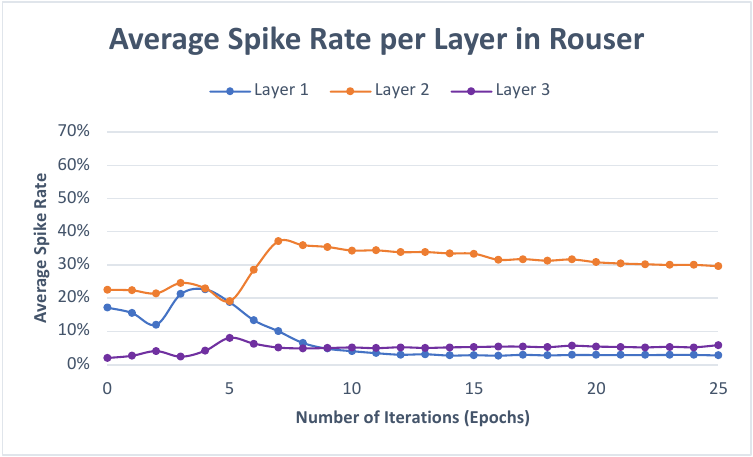}
\caption{}
\label{fig:spikes_b}
\end{subfigure}
\end{center}\vspace{-8pt}
\caption{Average spike rates.}\vspace{-8pt}
\label{spike-counts}
\end{figure}

We conducted a further investigation by examining the percentage of dead neurons in each layer during training. Fig.~\ref{dead-neurons} shows a lower count of ``dead neurons" in Rouser than the baseline. When the threshold level is high and neurons operate in a sub-threshold region, they have a near-zero gradient, and don't emit any spikes. This results in dead neurons, which are consequently unable to contribute to the learning process. By dynamically adjusting the threshold levels, Rouser enables a greater number of neurons to generate spikes and actively participate in error backpropagation. Additionally, in Fig.~\ref{spike-counts}, we present the average spike rate for NMNIST dataset. Initially, Rouser shows a higher average spike rate in layers 1 and 2 compared to the baseline. However, as training progresses, Rouser shows a reduction in spike rates. The average spike rate in layer 3 remains comparable, likely due to the use of the same loss function (MSE of the spike rate) for both training methods.

\subsection{Ablation Studies}
% We also conducted analysis studies on sub-optimal scenarios using the same loss function and initial set of hyperparameters (as in Table~\ref{hyperparameter}). Fig.~\ref{acc-all} shows the results. We observed a significant improvement in training latency (70\%) and accuracy (2\%) in Rouser, as shown in Fig.~\ref{nmnist:b} and ~\ref{nmnist:a}, respectively.

% \begin{figure}
%   \centering
%   \begin{subfigure}{0.35\textwidth}
%     \centering
%     \includegraphics[width=\textwidth]{Figures/Figs/NMNIST_a.pdf}
%     \caption{}
%     \label{nmnist:a}
%   \end{subfigure}
%   \hfill
%   \begin{subfigure}{0.35\textwidth}
%     \centering
%     \includegraphics[width=\textwidth]{Figures/Figs/NMNIST_b.pdf}
%     \caption{}
%     \label{nmnist:b}
%   \end{subfigure}
%   \caption{In a close comparison between Rouser and the Baseline, Rouser exhibits higher accuracy and lower latency for sub-optimal hyperparameters ($Th_{init}$, $\tau$, $s$) = (1.25, 3.75, 1.5) in (a), and ($Th_{init}$, $\tau$, $s$) = (0.25, 3, 3) in (b).}\vspace{-8pt}
%   \label{acc-all}
% \end{figure}

\begin{figure}[ht]
\begin{center}
\begin{subfigure}{0.35\textwidth}
    \centering
\includegraphics[width=\textwidth]{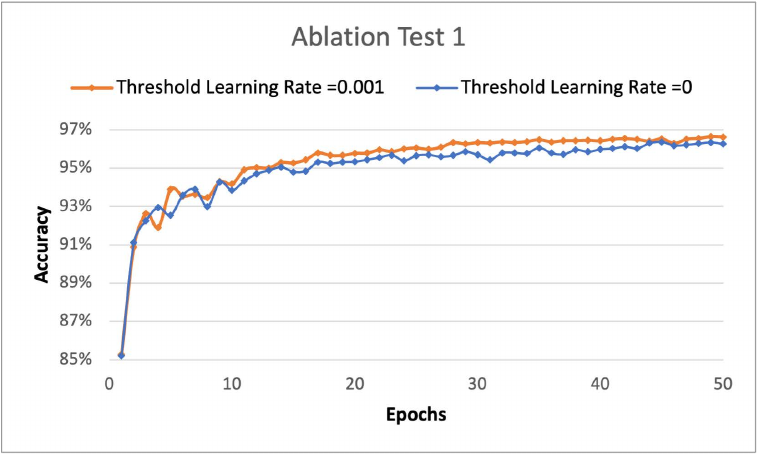}
\caption{($Th_{init}$, $\tau$, $s$) = (1.25, 3.75, 1.5)}
\label{fig:ablation_1}
\end{subfigure}
\hfill
\begin{subfigure}{0.35\textwidth}
\centering
\includegraphics[width=\textwidth]{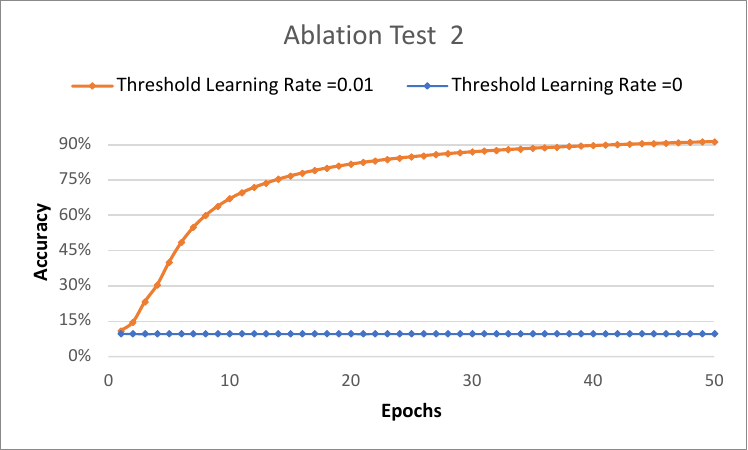}
\caption{($Th_{init}$, $\tau$, $s$) = (0.25, 3, 3)}
\label{fig:ablation_2}
\end{subfigure}
\end{center}\vspace{-8pt}
\caption{Ablation studies on NMNIST}\vspace{-8pt}
\label{fig:ablation}
\end{figure}

To determine whether the performance gains observed with Rouser are primarily due to threshold learning and not influenced by variations in experiemnts, we conducted ablation studies on Rouser. In these studies, we set the threshold learning rate to zero and ensured that identical hyperparameters and initial weights were used across experiments, while keeping all other variables such as network architecture unchanged. We then increased the threshold learning rates to assess its impact. We conducted these studies under two hyperparameter settings and observed that threshold learning rate of zero resulted in lower accuracy as shown in Fig.~\ref{fig:ablation_1}, and an inability to train in Fig.~\ref{fig:ablation_2}. 

\section{Conclusion}
\label{sec:conclusion}
In this work, we analyzed Rouser, a method for adaptively learning the threshold of neurons during SNN training, aimed at "rousing" dead neurons. By experimenting with three datasets—NMNIST, DVS128, and SHD—we demonstrated the effectiveness of this technique in enhancing network performance in terms of training latency and testing accuracy. We showed that Rouser's adaptive threshold learning consistently achieved high accuracy and robustness to variations in datasets and initial hyperparameters. Furthermore, Rouser significantly reduced the number of dead neurons during training while better regulating neuron spiking activity, resulting in faster training and improved accuracy.

% \section*{References}

\vspace{12pt}


\begin{thebibliography}{00}
\bibitem{retina}
M. Mahowald, "The Silicon Retina," in \emph{An Analog VLSI System for Stereoscopic Vision}, Springer US, Boston, MA, 1994, pp. 4--65.

\bibitem{cochlea}
L. Watts, D. A. Kerns, R. F. Lyon, and C. A. Mead, "Improved implementation of the silicon cochlea," \emph{IEEE Journal of Solid-State Circuits}, vol. 27, no. 5, pp. 692--700, 1992, doi: 10.1109/4.133156.

\bibitem{7047891}
Y. Nishitani, Y. Kaneko, and M. Ueda, "Supervised learning using spike-timing-dependent plasticity of memristive synapses," \emph{IEEE Transactions on Neural Networks and Learning Systems}, vol. 26, no. 12, pp. 2999--3008, Dec. 2015, doi: 10.1109/TNNLS.2015.2399491.

\bibitem{sheridan2017sparse}
P. M. Sheridan, F. Cai, C. Du, W. Ma, Z. Zhang, and W. D. Lu, "Sparse coding with memristor networks," \emph{Nature Nanotechnology}, vol. 12, no. 8, pp. 784--789, 2017.

\bibitem{strukov2008missing}
D. B. Strukov, G. S. Snider, D. R. Stewart, and R. S. Williams, "The missing memristor found," \emph{Nature}, vol. 453, no. 7191, pp. 80--83, 2008.

\bibitem{TrueNorth-IBM}
F. Akopyan \textit{et al.}, "TrueNorth: Design and tool flow of a 65 mW 1 million neuron programmable neurosynaptic chip," \emph{IEEE Trans. on CAD of Integrated Circuits and Systems}, vol. 34, no. 10, pp. 1537--1557, 2015. [Online]. Available: \url{http://dx.doi.org/10.1109/TCAD.2015.2474396}

\bibitem{hoppner2021spinnaker}
S. Höppner, Y. Yan, A. Dixius, S. Scholze, J. Partzsch, M. Stolba, F. Kelber, B. Vogginger, F. Neumärker, G. Ellguth, *et al.*, "The SpiNNaker 2 processing element architecture for hybrid digital neuromorphic computing," *arXiv preprint arXiv:2103.08392*, 2021.

\bibitem{davies2018loihi}
M. Davies \textit{et al.}, "Loihi: A neuromorphic manycore processor with on-chip learning," \emph{IEEE Micro}, vol. 38, no. 1, pp. 82--99, 2018.

\bibitem{AKIDA}
Akida Neural Processor, Brainchip, "Technology," \textit{[Online]}. Available: \url{https://www.brainchip.com/technology}. [Accessed: Apr. 2024].

\bibitem{GrAI}
GrAI Matter Labs, "Technology," \textit{[Online]}. Available: \url{https://www.graimatterlabs.ai/technology}. [Accessed: Apr. 2024].

% \bibitem{103389}
% T. DeWolf, P. Jaworski, and C. Eliasmith, "Nengo and low-power AI hardware for robust, embedded neurorobotics," *Front. Neurorobotics*, vol. 14, 2020. [Online]. Available: https://www.frontiersin.org/articles/10.3389/fnbot.2020.568359. doi: 10.3389/fnbot.2020.568359.

% \bibitem{9138762}
% G. Gallego, T. Delbrück, G. Orchard, C. Bartolozzi, B. Taba, A. Censi, S. Leutenegger, A. J. Davison, J. Conradt, K. Daniilidis, and D. Scaramuzza, "Event-based vision: A survey," *IEEE Trans. Pattern Anal. Mach. Intell.*, vol. 44, no. 1, pp. 154–180, Jan. 2022, doi: 10.1109/TPAMI.2020.3008413.

% \bibitem{1706816}
% M. Litzenberger, B. Kohn, A. N. Belbachir, N. Donath, G. Gritsch, H. Garn, C. Posch, and S. Schraml, "Estimation of vehicle speed based on asynchronous data from a silicon retina optical sensor," in *Proc. 2006 IEEE Int. Transportation Systems Conf.*, 2006, pp. 653–658, doi: 10.1109/ITSC.2006.1706816.

% \bibitem{sandamirskaya2022neuromorphic}
% Y. Sandamirskaya, M. Kaboli, J. Conradt, and T. Celikel, "Neuromorphic computing hardware and neural architectures for robotics," *Science Robotics*, vol. 7, no. 67, p. eabl8419, 2022.

% \bibitem{tang2019spiking}
% G. Tang, A. Shah, and K. P. Michmizos, "Spiking neural network on neuromorphic hardware for energy-efficient unidimensional SLAM," in *Proc. 2019 IEEE/RSJ Int. Conf. Intelligent Robots Systems (IROS)*, 2019, pp. 4176–4181.

\bibitem{kempter1999hebbian}
R. Kempter, W. Gerstner, and J. L. Van Hemmen, "Hebbian learning and spiking neurons," *Phys. Rev. E*, vol. 59, no. 4, p. 4498, 1999.

\bibitem{song2000competitive}
S. Song, K. D. Miller, and L. F. Abbott, "Competitive Hebbian learning through spike-timing-dependent synaptic plasticity," *Nat. Neurosci.*, vol. 3, no. 9, pp. 919–926, 2000.

\bibitem{diehl2015unsupervised}
P. U. Diehl and M. Cook, "Unsupervised learning of digit recognition using spike-timing-dependent plasticity," *Front. Comput. Neurosci.*, vol. 9, p. 99, 2015.

\bibitem{ferre2018unsupervised}
P. Ferré, F. Mamalet, and S. J. Thorpe, "Unsupervised feature learning with winner-takes-all based STDP," *Front. Comput. Neurosci.*, vol. 12, p. 24, 2018.

\bibitem{seung2003learning}
H. S. Seung, "Learning in spiking neural networks by reinforcement of stochastic synaptic transmission," *Neuron*, vol. 40, no. 6, pp. 1063–1073, 2003.

\bibitem{williams1992simple}
R. J. Williams, "Simple statistical gradient-following algorithms for connectionist reinforcement learning," *Mach. Learn.*, vol. 8, pp. 229–256, 1992.

\bibitem{cao2015spiking}
Y. Cao, Y. Chen, and D. Khosla, "Spiking deep convolutional neural networks for energy-efficient object recognition," *Int. J. Comput. Vis.*, vol. 113, pp. 54–66, 2015.

\bibitem{7280696}
P. U. Diehl, D. Neil, J. Binas, M. Cook, S.-C. Liu, and M. Pfeiffer, "Fast-classifying, high-accuracy spiking deep networks through weight and threshold balancing," in \emph{Proc. 2015 Int. Joint Conf. Neural Networks (IJCNN)}, 2015, pp. 1--8, doi: 10.1109/IJCNN.2015.7280696.

\bibitem{sengupta2019going}
A. Sengupta, Y. Ye, R. Wang, C. Liu, and K. Roy, "Going deeper in spiking neural networks: VGG and residual architectures," *Front. Neurosci.*, vol. 13, p. 95, 2019.

\bibitem{9556508}
N. Rathi and K. Roy, "DIET-SNN: A low-latency spiking neural network with direct input encoding and leakage and threshold optimization," *IEEE Trans. Neural Networks Learn. Syst.*, vol. 34, no. 6, pp. 3174–3182, Jun. 2023, doi: 10.1109/TNNLS.2021.3111897.

\bibitem{perez2021sparse}
N. Perez-Nieves and D. Goodman, "Sparse spiking gradient descent," \emph{Advances in Neural Information Processing Systems (NeurIPS)}, vol. 34, pp. 11795--11808, 2021.

\bibitem{bohte2000spikeprop}
S. M. Bohte, J. N. Kok, and J. A. La Poutré, "SpikeProp: backpropagation for networks of spiking neurons," in \emph{Proc. ESANN}, Bruges, 2000, pp. 419--424.

\bibitem{shrestha2018slayer}
S. B. Shrestha and G. Orchard, "Slayer: Spike layer error reassignment in time," \emph{Advances in Neural Information Processing Systems (NeurIPS)}, vol. 31, 2018.

\bibitem{superspike}
F. Zenke and S. Ganguli, "Superspike: Supervised learning in multilayer spiking neural networks," \emph{Neural Computation}, vol. 30, no. 6, pp. 1514--1541, 2018.

\bibitem{lee2016training}
J. H. Lee, T. Delbruck, and M. Pfeiffer, "Training deep spiking neural networks using backpropagation," *Front. Neurosci.*, vol. 10, p. 508, 2016.

\bibitem{kaiser2020synaptic}
J. Kaiser, H. Mostafa, and E. Neftci, "Synaptic plasticity dynamics for deep continuous local learning (DECOLLE)," *Front. Neurosci.*, vol. 14, p. 424, 2020.

\bibitem{8891809}
E. O. Neftci, H. Mostafa, and F. Zenke, "Surrogate Gradient Learning in Spiking Neural Networks: Bringing the Power of Gradient-Based Optimization to Spiking Neural Networks," \textit{IEEE Signal Processing Magazine}, vol. 36, no. 6, pp. 51-63, Nov. 2019, doi: 10.1109/MSP.2019.2931595.

\bibitem{zenke2021remarkable}
F. Zenke and T. P. Vogels, "The remarkable robustness of surrogate gradient learning for instilling complex function in spiking neural networks," *Neural Comput.*, vol. 33, no. 4, pp. 899–925, 2021.

% \bibitem{turrigiano2004homeostatic}
% G. G. Turrigiano and S. B. Nelson, "Homeostatic plasticity in the developing nervous system," *Nat. Rev. Neurosci.*, vol. 5, no. 2, pp. 97–107, Feb. 2004.

\bibitem{yin2020effective}
B. Yin, F. Corradi, and S. M. Bohté, "Effective and efficient computation with multiple-timescale spiking recurrent neural networks," in *Proc. Int. Conf. Neuromorphic Syst.*, 2020, pp. 1–8.

\bibitem{shaban2021adaptive}
A. Shaban, S. S. Bezugam, and M. Suri, "An adaptive threshold neuron for recurrent spiking neural networks with nanodevice hardware implementation," *Nature Commun.*, vol. 12, no. 1, p. 4234, Sep. 2021.

\bibitem{bellec2018long}
G. Bellec, D. Salaj, A. Subramoney, R. Legenstein, and W. Maass, "Long short-term memory and learning-to-learn in networks of spiking neurons," in *Advances in Neural Information Processing Systems (NeurIPS)*, vol. 31, 2018.

\bibitem{frequency}
D. Salaj, A. Subramoney, C. Kraisnikovic, G. Bellec, R. Legenstein, and W. Maass, "Spike frequency adaptation supports network computations on temporally dispersed information," \emph{eLife}, vol. 10, p. e65459, Jul. 2021, doi: 10.7554/eLife.65459. [Online]. Available: https://doi.org/10.7554/eLife.65459

\bibitem{annealing}
J. K. Eshraghian and W. D. Lu, "The fine line between dead neurons and sparsity in binarized spiking neural networks," \emph{CoRR}, vol. abs/2201.11915, 2022, [Online]. Available: https://arxiv.org/abs/2201.11915.

\bibitem{wang2022ltmd}
S. Wang, T. H. Cheng, and M. Lim, "LTMD: Learning improvement of spiking neural networks with learnable thresholding neurons and moderate dropout," *Adv. Neural Inf. Process. Syst. (NeurIPS)*, vol. 35, pp. 28350–28362, 2022.

% \bibitem{fang2021incorporating}
% W. Fang, Z. Yu, Y. Chen, T. Masquelier, T. Huang, and Y. Tian, "Incorporating learnable membrane time constant to enhance learning of spiking neural networks," in *Proc. IEEE/CVF Int. Conf. Comput. Vision*, 2021, pp. 2661–2671.

\bibitem{hasssan2024spiking}
Ahmed Hasssan, Jian Meng, and Jae-Sun Seo, ``Spiking Neural Network with Learnable Threshold for Event-based Classification and Object Detection,'' in \textit{Proceedings of the 2024 International Joint Conference on Neural Networks (IJCNN)}, pp. 1--8, IEEE, 2024.

\bibitem{LAVA}
Intel, "Lava Software Framework," \textit{[Online]}. Available: \url{https://lava-nc.org/}. [Accessed: Sep. 2023].

\bibitem{orchard}
G. Orchard, A. Jayawant, G. K. Cohen, and N. Thakor, "Converting static image datasets to spiking neuromorphic datasets using saccades," \textit{Frontiers in Neuroscience}, vol. 9, p. 437, 2015.

\bibitem{amir2017low}
A. Amir \emph{et al.}, "A low power, fully event-based gesture recognition system," in \emph{Proceedings of the IEEE Conference on Computer Vision and Pattern Recognition}, 2017, pp. 7243--7252.

\bibitem{cramer2020heidelberg}
B. Cramer, Y. Stradmann, J. Schemmel, and F. Zenke, "The Heidelberg spiking data sets for the systematic evaluation of spiking neural networks," *IEEE Trans. Neural Networks Learn. Syst.*, vol. 33, no. 7, pp. 2744–2757, Jul. 2020.









\end{thebibliography}
\end{document}